\documentclass[sigconf,natbib=true]{acmart}
\usepackage{ai-usage-card-fixed}
\usepackage{arydshln}
\usepackage{hyperref}
\usepackage{multirow}
\usepackage{geometry}
\usepackage{listings}
\usepackage{annotation}
\usepackage{xcolor}
\usepackage{amsmath} 
\usepackage{amsfonts}

\aiProjectName{MathRecSys}
\aiKeyApplication{Recommender System}

\aiContactName{Ankit Satpute}
\aiContactEmail{Ankit.Satpute@fiz-karlsruhe.de}
\aiContactAffiliation{FIZ Karlsruhe}

\aiModels{chatGPT, Claude}

\aiImprovingContent{Yes, but not used for generating new content}
\aiPerspectiveWork{Only self provided works}

    \aiFindingLiterature{No}

\aiRefactoringCode{Data handling}

\aiWhyUse{English is not the first language of authors so AI basically gives better text that would otherwise take many iterations to produce. However, original text is always produced by authors.}
\aiMitigateErrors{We double check the content by AI making sure it doesn't overstate anything we want to say.}
\aiMinimizeHarm{We check bias or prejudice and only include if it's backe by sources.}

\AtBeginDocument{%
  }

\copyrightyear{2026}
\acmYear{2026}
\setcopyright{cc}
\setcctype{by}
\acmConference[SIGIR '26]{Proceedings of the 49th International ACM SIGIR Conference on Research and Development in Information Retrieval}{July 20--24, 2026}{Melbourne, VIC, Australia}
\acmBooktitle{Proceedings of the 49th International ACM SIGIR Conference on Research and Development in Information Retrieval (SIGIR '26), July 20--24, 2026, Melbourne, VIC, Australia}
\acmDOI{10.1145/3805712.3809531}
\acmISBN{979-8-4007-2599-9/2026/07}

\newcommand{\oururlbase}{https://}
\newcommand{\reponame}{github.com/gipplab/MathAspectRecSys}
\newcommand{\repobase}{\oururlbase\reponame}
\newcommand{\repolink}[1]{\href{\repobase#1}{\url{\repobase#1}}}
\begin{document}

\title{Aspect-Aware Content-Based Recommendations for Mathematical Research Papers}

\author{Ankit Satpute}
\email{ankit.satpute@fiz-karlsruhe.de}
\orcid{0000-0003-3219-026X}
\affiliation{
  \institution{FIZ Karlsruhe}
  \streetaddress{Franklinstrasse 11}
  \city{Berlin}
  \country{Germany}
  \postcode{10587}
}

\author{Andr\'{e} Greiner-Petter}
\email{greinerpetter@gipplab.org}
\orcid{0000-0002-5828-5497}
\affiliation{
  \institution{University of Göttingen}
  \streetaddress{Wilhelmsplatz 1}
  \city{Göttingen}
  \country{Germany}
  \postcode{37073}
}

\author{Noah Gießing}
\email{noah.giessing@fiz-karlsruhe.de}
\orcid{0009-0006-5268-2519}
\affiliation{
  \institution{FIZ Karlsruhe}
  \streetaddress{Franklinstrasse 11}
  \city{Berlin}
  \country{Germany}
  \postcode{10587}
}

\author{Olaf Teschke}
\email{olaf.teschke@fiz-karlsruhe.de}
\orcid{0009-0003-4089-9647}
\affiliation{
  \institution{FIZ Karlsruhe}
  \streetaddress{Franklinstrasse 11}
  \city{Berlin}
  \country{Germany}
  \postcode{10587}
}

\author{Moritz Schubotz}
\email{moritz.schubotz@fiz-karlsruhe.de}
\orcid{0000-0001-7141-4997}
\affiliation{
  \institution{FIZ Karlsruhe}
\streetaddress{Franklinstrasse 11}
  \city{Berlin}
  \country{Germany}
  \postcode{10587}
}

\author{Akiko Aizawa}
\email{aizawa@nii.ac.jp}
\orcid{0000-0001-6544-5076}
\affiliation{
  \institution{National Institute of Informatics}
  \streetaddress{2 Chome-1-2 Hitotsubashi}
  \city{Tokyo}
  \country{Japan}
  \postcode{101-0003}
}

\author{Bela Gipp}
\email{gipp@uni-goettingen.de}
\orcid{0000-0001-6522-3019}
\affiliation{
  \institution{University of Göttingen}
  \streetaddress{Wilhelmsplatz 1}
  \city{Göttingen}
  \country{Germany}
  \postcode{37073}
}

\renewcommand{\shortauthors}{Satpute et al.}

\begin{abstract}
Content-based research paper recommendation (CbRPR) has seen advances in computer science and biomedicine, but remains unexplored for mathematics, where paper relatedness is more conceptual than explicit textual or citation-based similarity. 
Mathematics papers may be connected through shared proof techniques, logical implications, or natural generalizations, yet exhibit minimal textual or citation overlap, rendering existing CbRPR ineffective. 
To address this gap, we first conduct an expert-driven study characterizing mathematical recommendations, revealing that relevance is inherently \textit{aspect}-driven.
Grounded in this insight, we introduce GoldRiM (small, expert-annotated) and SilverRiM (large, automatically derived), the first datasets for \textit{aspect}-aware CbRPR in mathematics.
Recognizing that LLM embeddings of mathematical content alone yield suboptimal representation, we propose AchGNN, an \textit{aspect}-conditioned heterogeneous GNN that jointly models textual semantics, citation structure, and author lineage.
Across GoldRiM and SilverRiM, AchGNN consistently outperforms prior \textit{aspect}-based CbRPR methods, achieving substantial gains across all evaluated \textit{aspects}.
We conduct ablation studies to analyze the contributions of individual \textit{aspect} supervision, authorship lineage, and graph-structural signals to AchGNN’s performance.
To assess domain generality, we further evaluate AchGNN on the \textit{Papers with Code} dataset of machine learning publications, demonstrating that our \textit{aspect}-aware approach effectively transfers beyond mathematics.
We deploy our system on the MaRDI platform to help mathematicians with recommendations and release datasets and code publicly for reproducibility: \href{\repobase}{github.com/gipplab/MathAspectRecSys}.
\end{abstract}



\maketitle
\AddAnnotationRef

\section{Introduction}
\begin{figure}[t]
    \centering
    \includegraphics[width=\columnwidth,trim={0cm 7.5cm 15.1cm 0cm},clip]{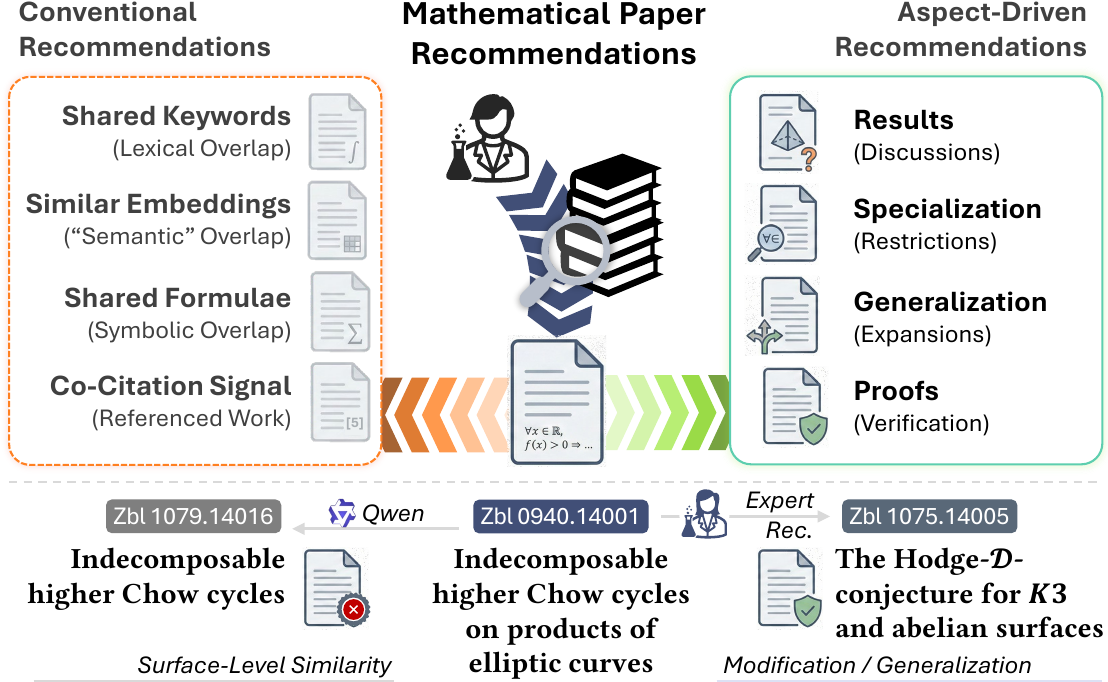}
    \caption{For a seed paper (center: \href{https://zbmath.org/0940.14001}{Zbl 0940.14001}), a CbRPR system (Qwen3-7B~\cite{qwen3embedding}) retrieves a surface-level similar paper (left: \href{https://zbmath.org/1079.14016}{Zbl 1079.14016}), whereas an expert mathematician recommends a paper (right: \href{https://zbmath.org/1075.14005}{Zbl 1075.14005}) that constitutes an important modification/generalization of the seed, despite minimal textual, semantic, or citation overlap.}
    \label{fig:math-content-motivation}
\end{figure}

Content-based Research Paper Recommendation (CbRPR) systems suggest scholarly work that is similar in content.
While such systems have seen progress in computer science (CS)~\cite{beel2016RSSurvey,kreutz2022scientific} and biomedicine (BM)~\cite{2022KartRSBio,2019BilmedRSFeng}, they remain unexplored for mathematics.
The gap is non-trivial as mathematical research differs fundamentally from CS and BM.
First, unlike experimental sciences, where overlap in terminologies, datasets, or methods is common, mathematical papers connect through abstraction, symbolic compression, and deductive reasoning~\cite{kiryu2023Math}.
Even citations in mathematics reference foundational concepts or proof techniques, rather than closely related prior work in the empirical sense~\cite{hulek2023mathAuthor}.
Second, candidates with high textual similarity (a basis effectively used in existing CbRPR~\cite{kreutz2022scientific,pinedo2025RSSurvey}) may be insufficient for capturing mathematical domain traits.
This is because two papers may be strongly related through, for instance, a dual formulation of a theorem or an extension of a proof strategy, while exhibiting minimal textual and citation overlap.
Figure~\ref{fig:math-content-motivation} demonstrates broader patterns: in mathematics, relevance is often determined by the domain-specific roles (e.g., generalization, restriction, or proof reuse), rather than by observable similarity signals.
Such domain-specific challenges, the scarcity of available datasets, and the lack of empirical evidence characterizing CbRPR in mathematics motivate our first research question (RQ1): \textbf{What constitutes content-based relevance in mathematical research papers?}

To address RQ1, we conducted the first expert-driven study of mathematical recommendations.
Our analysis in this work reveals that mathematical relevance is inherently \textit{aspect}-driven, encompassing conceptual, methodological, or structural connections (e.g., generalization of a theorem, dual formulation, shared proof technique).
Crucially, these math-specific \textit{aspect} categories are absent from existing CbRPR datasets.\footnote{\label{repo:DataSetComp}We provide additional materials, including an overview table of existing CbRPR datasets, alternative representations of results, aspect definitions, and more plots also in our repository: \href{\repobase}{github.com/gipplab/MathAspectRecSys}}
To empirically validate and operationalize this insight, we introduce two novel resources:
(1)~\textbf{GoldRiM}, a small high-quality test-dataset curated with expert recommendations; and
(2)~\textbf{SilverRiM}, a large-scale dataset derived from implicit recommendation signals in zbMATH Open\footnote{\label{zbmathlink}\url{https://zbmath.org/}} abstracts.
GoldRiM uncovers the structure of mathematical relevance, while SilverRiM provides the first large-scale benchmark for CbRPR in mathematics, with \textit{aspects} indicating \emph{why} papers are related in both datasets.

While leading CbRPR approaches rely on Large Language Model (LLM) embedding similarities~\cite{2025AspectCodes,pinedo2025RSSurvey,2025ZhangCiteIntentSurvey}, mathematical texts can cause LLM embeddings to underperform~\cite{fatima-2025-firma,24LLMMath}, suggesting that embeddings alone may be insufficient for mathematical CbRPR.
A mathematical CbRPR should prioritize domain-salient signals (e.g., keywords, venues, classification codes), but these are often coarse and inconsistently available, whereas authorship and citations offer a more robust and universal foundation.
Authorship also serves as a proxy for intellectual lineage, as mathematicians tend to work within narrowly defined subfields, reuse proof styles, and foundational results are popular with author names~\cite{hulek2023mathAuthor,2014Mathjournal}.
Hence, mathematical CbRPR should jointly model semantic similarity via citations and authorship lineage, enabling a system to prioritize works that are both semantically related and embedded within the same mathematical lineage.
This motivates our second research question (RQ2): 
\textbf{How does the joint modeling of textual and authorship lineage affect CbRPR in mathematics?}

To address RQ2, we require a modeling framework that can jointly capture semantic relatedness and intellectual lineage.
We leverage prior evidence that jointly modeling author–paper citation graphs through a Graph Neural Network (GNN) is effective for CbRPR~\cite{cummings2020AuthCit,yang2023heteroGNN,abdel22GNNLit}.
Since we find, via RQ1, that mathematical recommendations are inherently \textit{aspect}-driven, our model must also preserve \textit{aspect} specificity.
To this end, we formalize paper–paper recommendation using an \textsf{Aspect-conditioned Heterogeneous GNN (AchGNN)}.
The model operates on a heterogeneous graph with paper and author nodes and integrates semantic textual similarity, \textit{aspect} information, and authorship relations within a unified framework.

We evaluate AchGNN on GoldRiM and SilverRiM against several baselines, including fine-tuned LLM CbRPR~\cite{Ostendorff2022a}, a heterogeneous GNN~\cite{you2020graphCL}, and state-of-the-art LLM-embeddings~\cite{qwen3embedding,2025memtron}.
AchGNN consistently outperforms all competing baselines on GoldRiM and SilverRiM.
We further evaluate AchGNN on the Papers with Code (PwC) dataset~\cite{2020PapWitCodeData} to assess its applicability beyond heavily mathematics oriented benchmarks, where it demonstrates competitive performance.
The recommendations generated by AchGNN have been integrated into the Mathematical Research Data Initiative platform (MaRDI)\footnote{An example document with recommendations:\newline\url{https://portal.mardi4nfdi.de/wiki/Publication:3361952}} and are planned for incorporation into zbMATH Open in the near future.
Lastly, all our annotated datasets, the source code, and additional materials are publicly available.$^{\ref{repo:DataSetComp}}$

\section{Related Work}\label{sec:relatedWork}

\textit{Aspect-based CbRPR:} \textit{Aspect}-based CbRPR has been approached mainly through supervised classification and embedding-based retrieval~\cite{pinedo2025RSSurvey,2025ZhangCiteIntentSurvey}.
Supervised classification-based approaches~\cite{2020OstendorffWiki,cohan2019structural,kobayashi2018ACL} classify pairs of papers with respect to a predefined set of \textit{aspects} to produce recommendations.
However, their quadratic computational complexity makes them impractical for large-scale datasets, and they have shown limited ability to distinguish \textit{aspects} compared to embedding-based retrieval~\cite{2020OstendorffWiki,Ostendorff2022a}.
Early TF-IDF embedding-based retrieval methods~\cite{chakraborty2016ferosa,chan2018solventCSCW} were outperformed~\cite{Ostendorff2022a,2023aspectcse,2025AspectCodes} by fine-tuned models SciBERT~\cite{2019SciBERT} and SPECTER~\cite{2020-specter}.
Although fine-tuned embeddings consistently outperform general-purpose models, prior work has not addressed mathematical CbRPR.
The empirical success of existing \textit{aspect}-based CbRPR methods has largely been demonstrated on datasets$^{\ref{repo:DataSetComp}}$ drawn from CS and BM, which dominate current benchmarks and typically follow the Introduction-Method-Results-Discussion (IMRD) structure \cite{2022KartRSBio,1999CSSchematic}.
As a result, these methods are often based on LLM-derived document embeddings and are optimized for structurally regular documents.
Their effectiveness in domains with substantially different writing conventions remains underexplored~\cite{2023aspectcse}, particularly in mathematics, where text is semantically dense, highly symbolic, and often lacks explicit structural markers.

\begin{figure*}[ht]
    \centering
    \includegraphics[width=\textwidth,trim={0.5cm 7.3cm 7.5cm 5.4cm},clip]{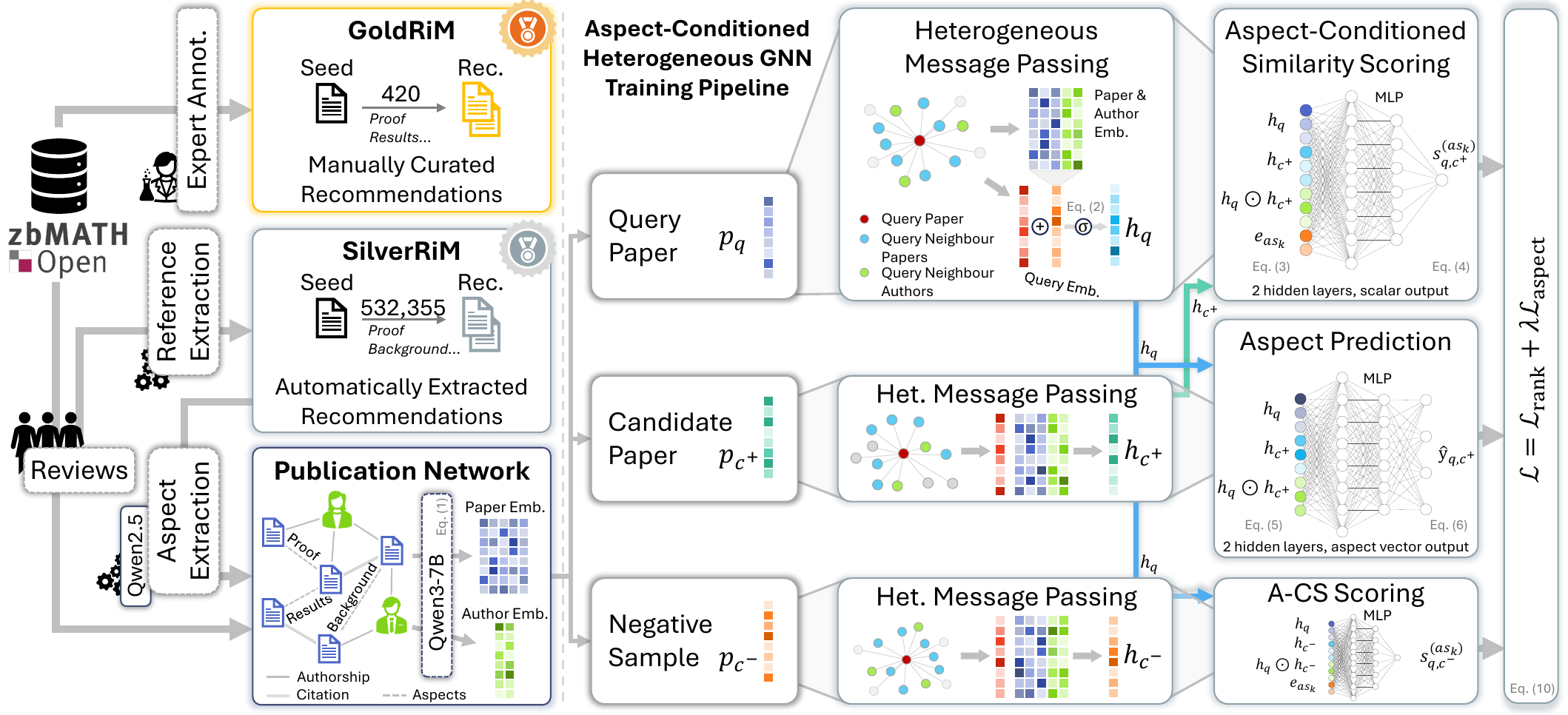}
    \caption{Architecture and learning pipeline of AchGNN.}
    \label{fig:AchGNN}
\end{figure*}

\textit{Heterogeneous GNNs in CbRPR:} Heterogeneous GNNs have shown consistent improvements in CbRPR over collaborative filtering and standalone embedding approaches by jointly modeling multiple scholarly entities, such as papers, authors, and venues~\cite{kreutz2022scientific,WANG2024HHaGPR,22DuanHetero}. 
These methods construct heterogeneous graphs and propagate information over citation, authorship, and venue relations to learn paper representations for recommendation via embedding similarity. 
While effective at leveraging structural signals, these models treat research papers as monolithic entities and do not produce \textit{aspects}-based recommendations.
Recent surveys confirm that although heterogeneous GNNs are widely adopted for CbRPR, \textit{aspect}-based CbRPR using GNNs has not been explored~\cite{beel2016RSSurvey,pinedo2025RSSurvey,23LIGNNSurvey}.
In contrast, \textit{aspect}-aware GNNs have been extensively studied in e-commerce recommendations, where user–item interaction graphs are enriched with \textit{aspect}-level information. 
In recent works, FigGNN~\cite{22WangFigNN} and MA-GNN~\cite{ZHANG2023Magnn} incorporate \textit{aspect}-conditioned user–item interactions to learn \textit{aspect}-specific embeddings that improve recommendation rankings.

\textit{Math Information Retrieval (MathIR):}
MathIR research primarily focuses on retrieving ranked similar mathematical content.
Here, we only mention works on the ArqMATH dataset~\cite{mansouri2022advancing}, the most recent dataset with human annotations on math content similarity.
A recent survey by Dadure et al.~\cite{malik2026review} provides a more comprehensive review of existing datasets and methods.
Mabowdor~\cite{23mabowdor}, a Math-Aware Best-of-Worlds Domain-Optimized Retriever, is the best-performing approach on the ArqMATH dataset. 
It combines an unsupervised structure search for similar math formulae with a dense retriever leveraging LLMs for text content similarity, integrating both into a combined score ranker.
Wang et al.~\cite{Wang2023MathIR} proposed CLFE, a contrastive learning framework that learns latent structures from formulas and generates embeddings.
Their work demonstrated that embedding-based k-nearest neighbor searches improve precision and recall.
A growing trend is the development of generalist LLMs capable of performing multiple tasks with minimal fine-tuning~\cite{behnamghader2024llmvec,malik2026review}, often outdoing domain-specific models.
These models are evaluated using benchmarks such as the Massive Text Embedding Benchmark (MTEB)~\cite{muennighoff-etal-2023-mteb}, which includes mathematics-specific sub-benchmarks.

\section{Methodology}
Motivated by success in e-commerce settings, we address the lack of an analogous formulation for mathematical CbRPR by integrating \textit{aspect}-aware modeling with authorship, treating shared authorship as a signal of conceptual lineage.
Figure~\ref{fig:AchGNN} provides an overview of our proposed CbRPR, consisting of two main stages: 
(1) grounding mathematical CbRPR, and 
(2) AchGNN: generating \textit{aspect}-based recommendations using a Heterogeneous GNN.
Because no dataset exists for mathematical CbRPR, the first stage focuses on analyzing how relevance is expressed in mathematical research and constructing datasets that operationalize this notion.
In the second stage, we design AchGNN, which integrates citation graph with authorship to model \textit{aspect}-based mathematical recommendations.

\subsection{Grounding Mathematical CbRPR}\label{sec:groundMathCbRPR}
Existing \textit{aspect}-based CbRPR datasets do not contain mathematical research papers~\cite{kreutz2022scientific,pinedo2025RSSurvey,beel2016RSSurvey}.
We therefore curate our own annotated dataset.
We choose zbMATH Open as the source because it excels in coverage, metadata quality, content availability, and accessibility via public APIs, compared to arXiv\footnote{\url{https://arxiv.org/}}, MathSciNet\footnote{\url{https://mathscinet.ams.org/mathscinet}}, or Google Scholar.
Most importantly, zbMATH Open's reviewing service offers a unique opportunity to curate expert recommendations.
zbMATH Open provides summaries (hereafter referred to as \textit{abstracts}) written by human expert reviewers.
These abstracts offer extensive historical coverage, spanning from 1763 to the present, and include many foundational works that are frequently cited.
The reviewers are domain experts, making their judgments a reliable source of relevance signals.
Reviewers of zbMATH Open are selected by the editor in charge based on their previous scientific contributions. 
They are usually among the foremost experts in the field; several Fields Medalists have contributed reviews to zbMATH Open.
This is critical because relevance in mathematical research is often implicit and conceptually complex, rendering expert judgment indispensable for establishing ground truth.
Utilizing zbMATH Open, we study and collect recommendations through two complementary datasets.
GoldRiM: emphasizes annotation quality and supports qualitative analysis of how mathematical recommendations are formed.
SilverRiM: emphasizes scale, enabling large-scale modeling and realistic quantitative evaluation.

\subsubsection{GoldRiM}
The goal of GoldRiM is to characterize mathematical relevance as perceived by domain experts, providing a qualitative foundation for mathematical CbRPR.
To this end, we collaborated with zbMATH Open to collect expert-provided recommendations, a common approach for constructing high-quality human-annotated datasets in CbRPR~\cite{pinedo2025RSSurvey,beel2016RSSurvey,kreutz2022scientific}.
Due to time and budget constraints, annotations were obtained from a single senior expert and active reviewer.
The reviewer had over 30 years of experience in curating mathematical literature, ensuring high-quality and internally consistent judgments.
We selected 80 seed documents, all drawn from the expert’s primary domain of expertise, Algebraic Geometry, to minimize domain mismatch and maximize annotation reliability.
For each seed document, the expert provided multiple recommendations, judged to be most relevant, yielding 420 seed-recommendation pairs (ranging from 3 to 11 per seed).
This collection constitutes a small but high-quality \textbf{Gold}-standard dataset of \textbf{R}ecommendations \textbf{i}n \textbf{M}athematics (\textbf{GoldRiM}).
We acknowledge likely bias and the limited scale of GoldRiM, but it is designed for qualitative and diagnostic analysis only.
Prior work has shown that even small, expert-curated datasets can yield meaningful and durable insights when the goal is to uncover underlying relevance structures rather than estimate population-level distributions~\cite{mysore2021csfcube,Jurgens2018ACLARC}.
In the following, we statistically analyze GoldRiM recommendations.

\begin{figure}[t]
    \centering
    \includegraphics[width=0.95\columnwidth]{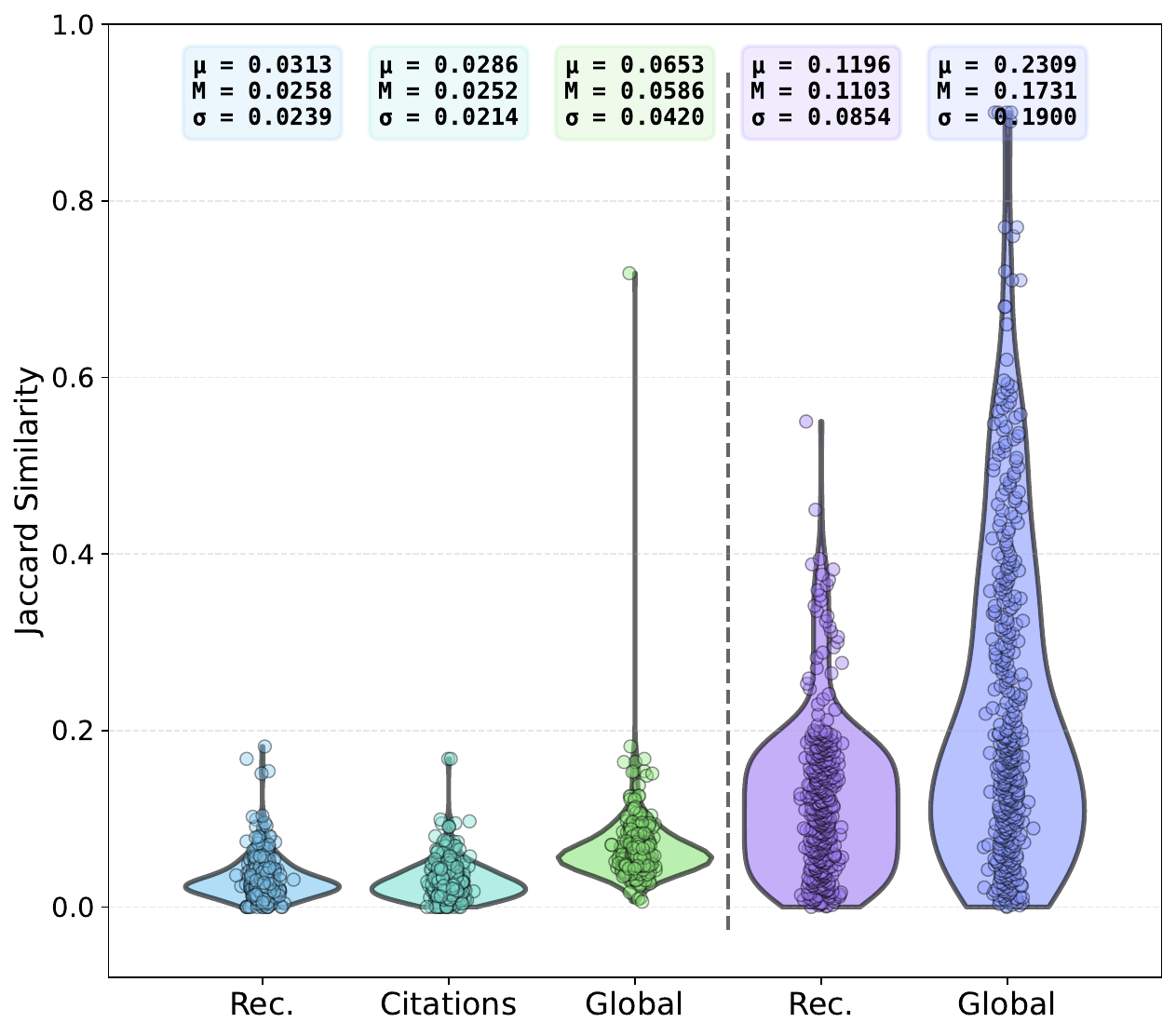}
    \caption{
    Distribution of 2-gram overlap between the GoldRiM seeds (left) and PwC seeds (right) and their respective recommendations (\textbf{Rec.}), citations (\textbf{Citations}), and all documents in the respective datasets (\textbf{Global}). PwC does not provide citation data; hence, the citation distribution is unavailable. Citations and global entries are capped to 420 top-scoring articles. $\mu$: mean, $M$: median, and $\sigma$: standard deviation.}
    \label{fig:violin}
\end{figure}

\textbf{Lexical overlap:} Naively, relevance has been measured via surface-level lexical overlap between seeds and recommendations (see TF-IDF approaches in Section~\ref{sec:relatedWork}).
In mathematics, however, relevance is often rooted in deep connections between mathematical concepts that may or may not be readily apparent from lexical overlaps.
Figure~\ref{fig:violin} shows the Jaccard similarity based on 2-grams\footnote{We further analyzed 1- and 3-gram overlap. However, 1-grams produced higher scores but exhibited the same patterns as 2-grams, and 3-gram overlap yielded near-zero overlap. We therefore discuss only 2-gram analysis here.} overlaps in GoldRiM and PwC.
Across GoldRiM sets, similarity values are uniformly low compared to PwC sets.
Specifically, GoldRiM recommendations exhibit a mean similarity of just 0.031, with a tightly concentrated distribution near zero, demonstrating that expert recommendations are not driven by lexical overlap.
Interestingly, citation-based pairs show an even lower overlap, indicating that citation adjacency does not correspond to lexical similarity in mathematical literature.
In PwC, the overlap and similarity score varies significantly more depending on the selected seeds, with some seeds showing distributions similar to those in zbMATH Open.
However, the other end of the spectrum, shown in Figure~\ref{fig:violin}, exhibits significantly more lexical overlap.

\begin{figure}[t]
    \centering
    \includegraphics[width=0.95\columnwidth]{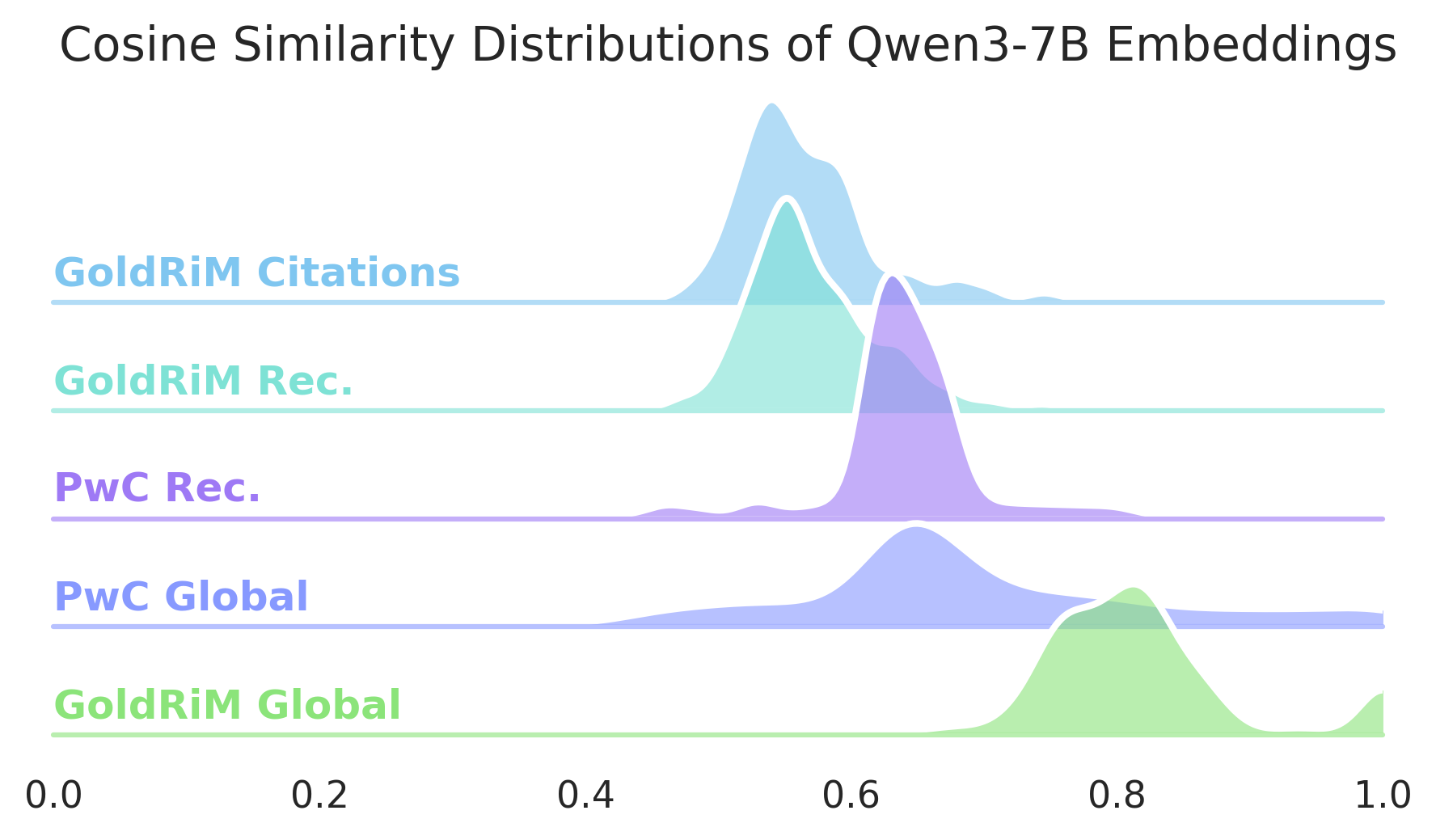}
    \caption{
    Distribution of 420 Qwen3-7B embedding similarities in GoldRiM and PwC sorted by their respective mean values between the seeds and their recommendations (Rec.), citations (Citations), and all documents from the representative datasets (Global). PwC does not contain citation data; hence, the citation distribution is unavailable.}
    \label{fig:embeddings_joyplot}
\end{figure}

\textbf{Embeddings analysis:} 
Semantic similarity analysis via embeddings has typically yielded promising results in many search or CbRPR tasks~\cite{Ostendorff2022a,2025AspectCodes}.
Figure~\ref{fig:embeddings_joyplot} illustrates the semantic distances between seeds and recommendations based on Qwen3-7B~\cite{qwen3embedding} embeddings in GoldRiM and PwC.
Again, GoldRiM citations and recommendations exhibit the lowest similarity.
However, the difference to recommendations in PwC is not as pronounced as for the lexical comparison above.
More crucially, however, is the clear overlap between similarity scores of PwC recommendations and the most similar articles in the PwC dataset.
This suggests that highly similar embeddings correlate with relevance in PwC.
In contrast, this correlation is not evident in GoldRiM.
In fact, recommendations and citations have significantly lower similarity scores than those available within zbMATH Open (Global), further underscoring that embeddings alone are insufficient for capturing expert-level recommendations in mathematics.

\textbf{Citation-proximity analysis:}
Both analyses above indicate a correlation between citations and recommendations within GoldRiM.
However, we find that 23\% of GoldRiM recommendations have no citation path between the seed and the recommended paper.
Among the remaining pairs, 43\% are connected by a single citation edge and only 3\% by two edges.
Such short paths are not discriminative, as many non-recommended papers are also reachable within one or two jumps.
For the remaining 31\% of recommendations, the shortest citation path exceeds three edges, requiring traversal through multiple intermediary documents.
Bibliographic coupling achieves a recall of 34\%, while co-citation achieves 35\%.
These results indicate that citation structures are inherently important for recommendations, but capture only a rather limited subset of expert judgments.
As such, solely relying on citation networks alone likely results in an insufficient pool of recommendations.
After consulting with the zbMATH Open expert, taking authorship lineage into account can be considered a natural extension of the citation network.
Mathematicians tend to reuse proof styles and foundational results or methodologies are often referred to by the author's names rather than specialized descriptive terms~\cite{hulek2023mathAuthor,2014Mathjournal}.

\textbf{Aspect-based similarity:}
The inability to rely solely on text similarity or citation-based signals indicates that mathematical relevance is governed by implicit relationships.
This motivated a qualitative analysis of how expert recommendations are conceptually related to one another.
Together with the expert, we manually analyzed all 420 recommendation pairs and assigned labels to describe the underlying relationships.
This analysis revealed that recommendations are connected through distinct \textit{aspects}, each capturing a specific mode of mathematical reasoning.
The initial labeling yielded 66 \textit{aspects} (detailed list is available online).$^{\ref{repo:DataSetComp}}$
Many labels, however, reflected surface-level linguistic variation rather than fundamentally different relationships.
For example, labels such as \textit{restriction}, \textit{reduction}, and \textit{limiting case} differ lexically but express the same conceptual relationship.
Furthermore, \textit{aspect}-based retrieval requires \textit{aspect} spaces small enough to allow statistically meaningful evaluation and tractable supervision.
In order to enable comparability with community standards, we consolidated the original labels into four high-level \textit{aspects}, \textit{Specialization / Restriction} (size: 32.14\%), \textit{Modification / Generalization} (40\%), \textit{Results} (17.86\%), and \textit{Prove / Cases} (10\%) through iterative discussion with a domain expert (detailed descriptions of each category are available in our repository).$^{\ref{repo:DataSetComp}}$
This abstraction preserves conceptual distinctions while enabling generalization and interpretability.

\subsubsection{SilverRiM}
\begin{figure}[t]
    \centering
    \includegraphics[width=0.45\textwidth,trim={0.7cm 1.6cm 0.7cm 0},clip]{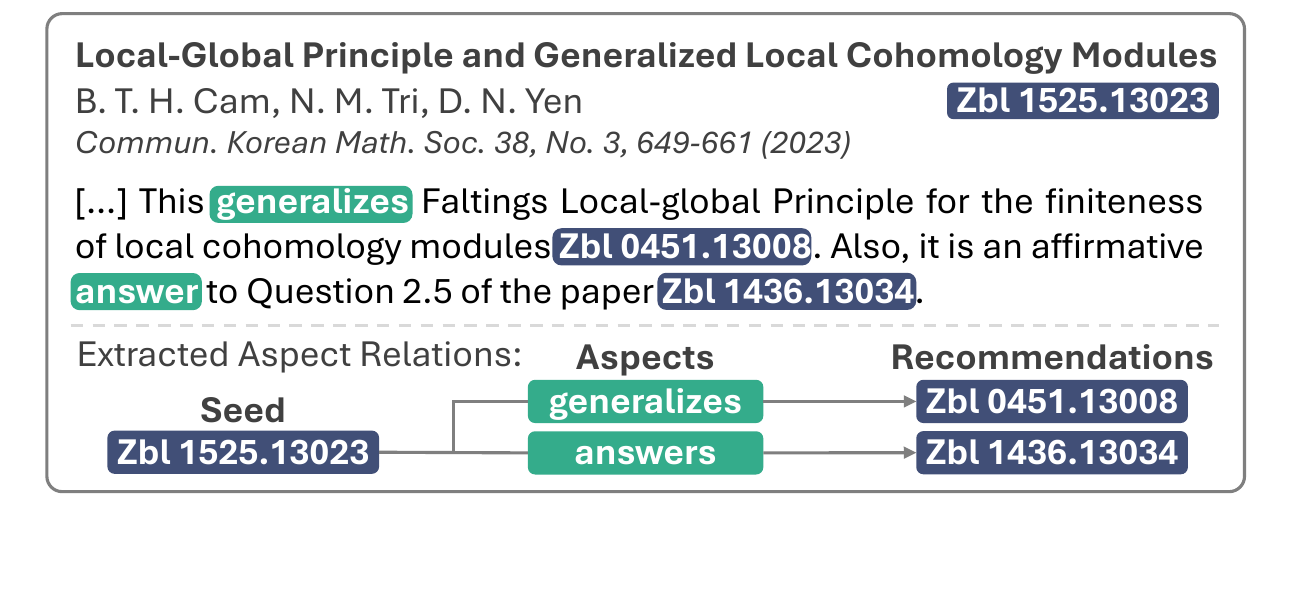}
    \caption[An example seed from zbMATH Open]{
        Example of \textit{aspect} extractions.
    }
    \label{fig:exampleSILVER}
\end{figure}

In essence, the GoldRiM analysis answers RQ1.
GoldRiM shows that relevance does not seem to correlate with lexical or embedding similarities but can be partially derived from citations and \textit{aspect} connections.
GoldRiM's inherent flaws are its limited size and bias from a single annotator.
It necessitates a scalable dataset that covers a broader area of zbMATH Open and a greater variety in generating \textit{aspects} and relevance.
During the construction of GoldRiM, we observed that many of the manually identified \textit{aspects} were explicitly articulated in zbMATH Open reviewer-written abstracts, often through references to prior work (see Figure~\ref{fig:exampleSILVER}). 
Considering zbMATH Open's reviewers are often highly decorated field experts themselves, we can consider such in-abstract references as recommendations.
In combination, this allows us to generate comparable data to GoldRiM at scale.

As of October 2025, zbMATH Open contains approximately 312K abstracts with 786K in-abstract references by over 13k different reviewers, providing a natural basis for large-scale extraction of mathematical recommendations and \textit{aspect}-labels.
Prior work shows that few-shot LLM-based phrase extraction can approximate human annotations with competitive quality in citation-related tasks~\cite{tan-etal-2024-large,qin-etal-2023-chatgpt}.
Therefore, we use Qwen2.5-14B~\cite{qwen2025qwen25technicalreport}, which performs strongly in citation-intent prediction and extraction~\cite{koloveas2025can}, in order to extract \textit{aspect}-labels.
Given an abstract and its in-abstract references, the model is prompted to extract short descriptive phrases characterizing the relationship between the seed paper and each referenced work.
After filtering malformed outputs and missing contexts, we obtain 212K seed papers, 532K recommendation pairs, and 7,326 distinct relationship labels.
This collection constitutes a large-scale \textbf{Silver}-standard dataset of \textbf{R}ecommendations \textbf{i}n \textbf{M}athematics (\textbf{SilverRiM}).

\begin{figure}[!t]
    \centering
    \includegraphics[width=0.99\columnwidth]{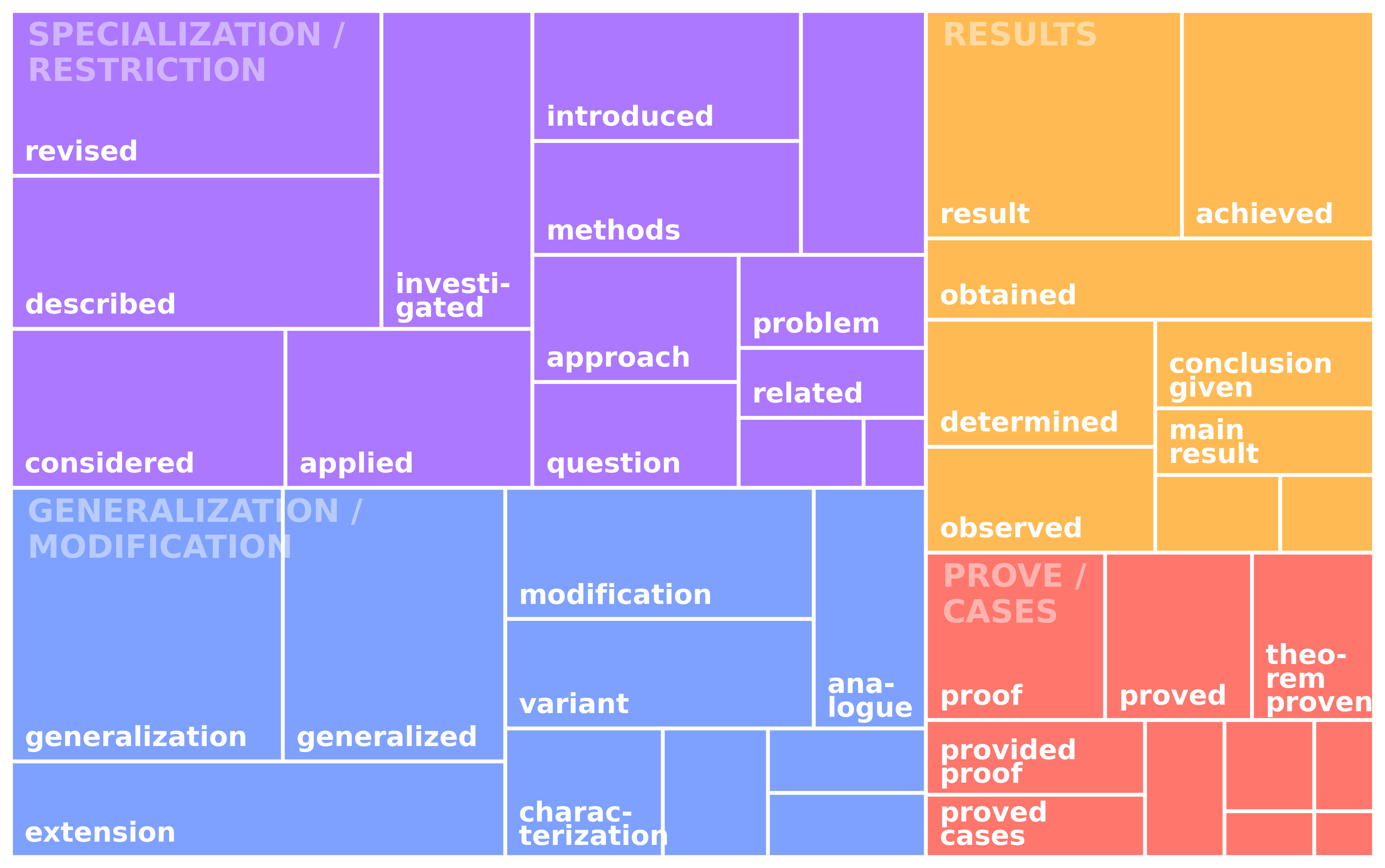}
    \caption{
    Fine-grained \textit{aspect} labels on SilverRiM.%
    }
    \label{fig:aspect_labels}
\end{figure}

The extracted relationship labels capture fine-grained linguistic expressions used by reviewers, but they are inherently noisy and highly sparse.
Treating them as independent ground-truth \textit{aspects} hinders generalization, evaluation, and interpretability.
In \textit{aspect}-based CbRPR, aspects encode more semantically informative relationships than their exact textual wording, a pattern consistently observed in prior domain-specific CbRPR work~\cite{Ostendorff2022a,2025AspectCodes,pinedo2025RSSurvey}.
Hence, we align the SilverRiM aspect-labels with the established GoldRiM ontology of four categories using k-means clustering.
Manual inspection of representative labels from each cluster confirms semantic alignment with the corresponding expert-defined \textit{aspects}.
Figure~\ref{fig:aspect_labels} shows the most frequently \textit{aspect}-labels of SilverRiM in their associated cluster.
Although automatically derived, the dataset resembles a very similar distribution as in GoldRiM, with \textit{Specialization/Restriction} (37.83\%) and \textit{Generalization/Modification} (29.30\%), followed by \textit{Results} (21.04\%) and \textit{Prove/Cases} (11.83\%).
Table~\ref{tab:dataset_characteristics} provides an overview of the curated datasets in comparison with PwC.

\begin{table}[t]
    \centering
    \caption{Statistics of the evaluation datasets. \textit{Aspect}-labels refers to the number of fine-
    grained generated labels.}
    \label{tab:dataset_characteristics}
    \begin{tabular}{lrrr}
        \hline
        \textbf{Characteristic} & \textbf{GoldRiM} & \textbf{SilverRiM} & \textbf{PwC} \\
        \hline
        Total Seeds & 80 & 212,382 & 157,606\\
        Rec. Pairs & 420 & 532,355 & 1,227,058\\
        Unique \textit{Aspects} & 4 & 4 & 3\\
        \textit{Aspect}-labels & 66 & 7,326 & 3,952\\
        Avg. Rec. per Seed & 5.25 & 2.50 & 7.78\\
        \hline
    \end{tabular}
\end{table}

A comparison with existing \textit{aspect}-based CbRPR datasets, which largely focus on CS and BM literature, reveals limited overlap and substantial domain-specific divergence. 
While the \textit{Results} \textit{aspect} is shared, other \textit{aspects} in our dataset are mathematics-specific and capture relations not represented in prior datasets. 
This highlights mathematics as a distinct and previously underrepresented domain for CbRPR, reflecting fundamentally different retrieval and recommendation needs. 
Our dataset, therefore, fills an important gap and enables the study of \textit{aspect}-aware recommendations in a new and challenging domain.

\subsection{Aspect-Conditioned Heterogeneous GNN}\label{sec_aspectCond}
Prior work~\cite{2022GraphCitRec,kreutz2022scientific} has shown that heterogeneous GNNs effectively exploit citation and authorship structure for CbRPR; however, existing formulations learn a single, \textit{aspect}-agnostic representation per paper (see Section~\ref{sec:relatedWork}).
Motivated by \textit{aspect}-conditioned graph training in e-commerce recommendations~\cite{22WangFigNN,ZHANG2023Magnn}, we adapt \textit{aspect}-aware modeling to mathematical CbRPR by integrating \textit{aspect} conditioning into a heterogeneous citation–authorship graph.
Authorship serves as a domain-salient signal of conceptual lineage, while \textit{aspect} conditioning enables fine-grained distinctions among relevance relations during training.
This results in an \textbf{A}spect-\textbf{c}onditioned \textbf{h}eterogeneous \textbf{GNN} (AchGNN), which combines heterogeneous message passing with \textit{aspect}-conditioned scoring to support \textit{aspect}-based CbRPR in mathematics.

\subsubsection{Graph Initialization}
We define a heterogeneous undirected graph $\mathcal{G} = (\mathcal{V}, \mathcal{E})$ with two node types: papers $\mathcal{P}$ and authors $\mathcal{A}$, such that $\mathcal{V} = \mathcal{P} \cup \mathcal{A}$. 
The edge set captures three complementary relations: \textbf{Aspect-labeled edges} $(p_i, p_j, as_k) \in \mathcal{E}_{as}$, indicating that papers $p_i$ and $p_j$ are similar with \textit{aspect} $as_k \in \mathcal{AS}$, \textbf{Citations} $(p_i, p_j) \in \mathcal{E}_{c}$, and \textbf{Authorship edges} $(a_m, p_i) \in \mathcal{E}_a$, connecting authors to their papers and encoding intellectual lineage with $\mathcal{E} = \mathcal{E}_{as} \cup \mathcal{E}_a \cup \mathcal{E}_c$.
This construction enables information flow not only between textually linked papers but also across papers connected through shared authors.
Paper nodes are initialized using titles and abstract embeddings (with in-abstract citations removed).
Author nodes are initialized using Author-name embeddings.
\begin{equation}
h_i^{(0)} = f_{\text{text}}(p_i) \in \mathbb{R}^{d},  h_l^{(0)} = f_{\text{text}}(a_m) \in \mathbb{R}^{d} 
\end{equation}
where $f_{\text{text}}$ is an embedding obtained using Qwen3-7B, as it yields the strongest performance among embedding-based base model baselines on GoldRiM and SilverRiM (see Section~\ref{sec:evaluatedmodels}).
Following prior heterogeneous GNN approaches that incorporate author nodes~\cite{cummings2020AuthCit,yang2023heteroGNN}, we use the same embedding model to encode both papers and author names, ensuring a shared representation space and consistent embedding dimensionality across node types.
Author representations are refined via message passing along authorship edges, enabling authors to inherit semantic signals from their associated papers. 
In this way, authorship acts as a structural proxy for conceptual lineage.

\subsubsection{Heterogeneous Message Passing}
We employ a GraphSAGE-inspired heterogeneous GNN Message passing~\cite{2017GraphSage} given its scalability and empirical robustness in scholarly graphs~\cite{zhong2023hierarchical}.
We stack $L$ GNN layers and use fixed neighborhood sampling to balance coverage and noise that has been observed to be optimal in heterogeneous scholarly graphs~\cite{you2020graphCL,2017GraphSage}.
Let $h_v^{(l)}$ denote the aggregated embedding of node $v$ at layer $l$. 
Node updates are computed as:
\begin{equation}
h_v^{(l+1)} = \sigma \Bigg(W_0^{(l)} h_v^{(l)} + \sum_{r \in \mathcal{R}}\sum_{u \in \mathcal{N}_r(v)}\frac{1}{|\mathcal{N}_r(v)|} W_r^{(l)} h_u^{(l)}\Bigg),
\end{equation}
where $\mathcal{R}$ denotes the relation types $\{c, a\}$ (citation or authorship), $\mathcal{N}_r(v)$ the neighbors of $v$ under relation $r$, and $W_r^{(l)}$ relation-specific transformation matrices. 
The resulting embeddings encode both connected papers via citations and authorship lineage.

\subsubsection{Aspect-Conditioned Similarity Scoring}
\textit{Aspect} conditioning is applied at the scoring level rather than during message passing to avoid entangling multiple \textit{aspects} within node embeddings.
We combine the paper-to-node transformation proposed in GraphCL \cite{you2020graphCL} with \textit{aspect}-conditioning used in FigGNN \cite{22WangFigNN} and MA-GNN \cite{ZHANG2023Magnn}.
Given a query paper $p_q$, a candidate paper $p_c$, and an \textit{aspect} $as_k \in \mathcal{AS}$, we construct an \textit{aspect}-conditioned interaction vector:
\begin{equation}
    z_{q,c}^{(as_k)} = \big[\, h_q \;\| \; h_c \;\| \; h_q \odot h_c \;\| \; e_{as_k} \,\big],
\end{equation}
where $h_q, h_c \in \mathbb{R}^d$ are the final GNN embeddings of the query and candidate papers, $\odot$ denotes element-wise multiplication, and $e_{as_k} \in \mathbb{R}^{d_{as}}$ is a learnable embedding associated with \textit{aspect} $as_k$.
This design explicitly exposes both symmetric interactions ($h_q \odot h_c$) and \textit{aspect} identity, allowing the scoring function to learn \textit{aspect}-specific similarity patterns.
The interaction vector is mapped to a scalar relevance score by a standard Multi-Layer Perceptron (MLP) $f_{\text{score}}$: 
\begin{equation}
    s_{q,c}^{(as_k)} = f_{\text{score}}\!\left(z_{q,c}^{(as_k)}\right).
\end{equation}
\subsubsection{Aspect Prediction Objective}
We introduce an auxiliary \textit{aspect}-prediction task to encourage \textit{aspect}-discriminative representations.
For a paper pair $(p_q, p_c)$, we construct an aspect-prediction vector:
\begin{equation}
    z_{q,c} =
    \big[\, h_q \;\| \; h_c \;\| \; h_q \odot h_c \,\big].
\end{equation}
The interaction vector is passed through an aspect-classification network:
\begin{equation}
    \hat{\mathbf{y}}_{q,c} = f_{\text{aspect}}\!\left(z_{q,c}\right),
\end{equation}
where $\hat{\mathbf{y}}_{q,c} \in \mathbb{R}^{|\mathcal{AS}|}$ denotes the predicted aspect logits.
The aspect classifier is implemented as an MLP with an output dimension equal to the number of aspects:
\begin{equation}
    f_{\text{aspect}}(z)
    = W_a \, \sigma\!\left(W_z z + b_z\right) + b_a.
\end{equation}
The auxiliary loss is the standard cross-entropy loss:
\begin{equation}
    \mathcal{L}_{\text{aspect}}
    = \mathrm{CE}\!\left(
        \mathrm{softmax}\!\left(\hat{\mathbf{y}}_{q,c}\right),
        as_k
    \right).
\end{equation}
Aspect prediction is formulated as a multi-class classification problem over observed positive paper pairs. Cross-entropy loss provides implicit negative supervision through competition among aspect classes, and negative paper pairs are not used, as their aspects are undefined.

\subsubsection{Learning Objectives}
We jointly optimize AchGNN for (i) aspect-conditioned separation and (ii) aspect prediction.
For aspect-conditioned separation, we adopt a pairwise Bayesian Personalized Ranking (BPR) loss~\cite{2009BPRLoss}:
\begin{equation}
    \mathcal{L}_{\text{rank}} =
    - \log \sigma \big( s_{q,c^+}^{(e_{as_k})} - s_{q,c^-}^{(e_{as_k})} \big).
\end{equation}
encouraging higher scores for aspect-consistent positive edges ($c^+$) and lower scores for negative edges ($c^-$).
The classification loss is standard cross-entropy over aspects. 
The final objective is:
\begin{equation}
    \mathcal{L}
    = \mathcal{L}_{\text{rank}} + \lambda \, \mathcal{L}_{\text{aspect}},
\end{equation}
where $\lambda$ controls the contribution of aspect supervision. 
This joint formulation regularizes ranking by enforcing aspect-discriminative representations.

\subsubsection{Inference.}
At inference time, AchGNN computes contextual embeddings for all papers, evaluates aspect-conditioned scores $s_{q,c}^{(as)}$, and ranks candidates accordingly:
\begin{equation}
    \text{TopN}^{(as)}(p_q) = \mathrm{argsort}_{c}\; s_{q,c}^{(as)}.
\end{equation}
This yields aspect-specific recommendation lists that reflect both relational proximity and mathematical lineage.

\section{Experiments}\label{sec:experiments}
We conduct experiments on GoldRiM, SilverRiM, and PwC~\cite{2020PapWitCodeData} to address RQ2. 
PwC consists of machine learning papers annotated with structured aspects: \textit{task}, \textit{method}, and \textit{dataset}, and enables evaluation beyond the mathematics domain.
Furthermore, PwC has been widely used as a benchmark for aspect-based CbRPR in the past~\cite{Ostendorff2022a,2025AspectCodes,pinedo2025RSSurvey}.

\begin{table*}[t]
\centering
\caption{Overall results for $k = 10$ retrieved documents in \textbf{GoldRiM} and \textbf{SilverRiM}. Precision (P), recall (R), and mean reciprocal rank (MRR) are reported.}
\small
\begin{tabular}{lcccccccccccccccccccc}
\toprule
Aspects $\longrightarrow$ & \multicolumn{3}{c}{General} & \multicolumn{3}{c}{Specialization /} & \multicolumn{3}{c}{Results} & \multicolumn{3}{c}{Prove / Cases} & \multicolumn{3}{c}{Modification /}\\
& & & & \multicolumn{3}{c}{Restriction} &  & &  &  &  &  & \multicolumn{3}{c}{Generalization}\\
\cmidrule(lr){2-4} \cmidrule(lr){5-7} \cmidrule(lr){8-10} \cmidrule(lr){11-13} \cmidrule(lr){14-16}
Methods $\downarrow$ & P & R & MRR & P & R & MRR & P & R & MRR & P & R & MRR & P & R & MRR\\
\midrule
\multicolumn{3}{l}{\textbf{Dataset: GoldRiM}} & & & & & & & & & & & & \\
\hdashline
Approach0 & 0.050 & 0.098 & 0.191 & 0.043 & 0.085 & 0.163 & 0.035 & 0.070 & 0.135 & 0.030 & 0.060 & 0.118 & 0.020 & 0.042 & 0.082 \\
Mabowdor & 0.032 & 0.064 & 0.102 & 0.027 & 0.056 & 0.089 & 0.021 & 0.043 & 0.071 & 0.016 & 0.033 & 0.054 & 0.012 & 0.026 & 0.043 \\ 
DPR & 0.035 & 0.074 & 0.125 & 0.030 & 0.065 & 0.108 & 0.024 & 0.052 & 0.089 & 0.018 & 0.039 & 0.070 & 0.014 & 0.031 & 0.055 \\  
SPECTER & 0.023 & 0.050 & 0.111 & 0.017 & 0.038 & 0.083 & 0.012 & 0.029 & 0.061 & 0.008 & 0.021 & 0.044 & 0.006 & 0.016 & 0.031 \\
SciBERT     & 0.007 & 0.016 & 0.035 & 0.005 & 0.012 & 0.026 & 0.004 & 0.009 & 0.019 & 0.003 & 0.007 & 0.014 & 0.002 & 0.005 & 0.010 \\
SciBERT-FT  & 0.075 & 0.133 & 0.261 & 0.061 & 0.112 & 0.228 & 0.047 & 0.091 & 0.193 & 0.035 & 0.072 & 0.161 & 0.026 & 0.054  & 0.129 \\
Qwen2-7B    & 0.077 & 0.156 & 0.391 & 0.062 & 0.131 & 0.332 & 0.048 & 0.102 & 0.274 & 0.035 & 0.078 & 0.221 & 0.026 & 0.059 & 0.176 \\
Qwen3-7B    & 0.106 & 0.213 & 0.413 & 0.089 & 0.184 & 0.361 & 0.071 & 0.148 & 0.302 & 0.054 & 0.116 & 0.248 & 0.041 & 0.089 & 0.201 \\
Qwen3-7B-FT   & 0.122 & 0.245 & \underline{0.475} & 0.102 & 0.212 & 0.415 & 0.082 & 0.170 & 0.347 & 0.062 & 0.133 & 0.285 & 0.047 & 0.102 & 0.231 \\
Memtron     & 0.091 & 0.187 & 0.343 & 0.075 & 0.159 & 0.297 & 0.059 & 0.128 & 0.243 & 0.044 & 0.101 & 0.196 & 0.033 & 0.077 & 0.156 \\
GraphCL     & \underline{0.271} & \underline{0.352} & 0.471 & \underline{0.238} & \underline{0.319} & \underline{0.431} & \underline{0.203} & \underline{0.284} & \underline{0.392} & \underline{0.172} & \underline{0.251} & \underline{0.351} & \underline{0.143} & \underline{0.217} & \underline{0.309} \\
AchGNN & \textbf{0.341} & \textbf{0.425} & \textbf{0.593} & \textbf{0.312} & \textbf{0.398} & \textbf{0.556} & \textbf{0.279} & \textbf{0.361} & \textbf{0.512} & \textbf{0.243} & \textbf{0.326} & \textbf{0.468} & \textbf{0.211} & \textbf{0.289} & \textbf{0.423} \\
\hdashline
AchGNN\_noMath & 0.324 & 0.403 & 0.566 & 0.296 & 0.379 & 0.533 & 0.266 & 0.345 & 0.491 & 0.232 & 0.311 & 0.450 & 0.201 & 0.276 & 0.407 \\
\midrule
\multicolumn{3}{l}{\textbf{Dataset: SilverRiM}} & & & & & & & & & & & &  \\
\hdashline
Approach0 & 0.062 & 0.460 & 0.325 & 0.064 & 0.475 & 0.340 & 0.058 & 0.430 & 0.300 & 0.060 & 0.445 & 0.315 & 0.055 & 0.420 & 0.295 \\
Mabowdor & 0.048 & 0.370 & 0.255 & 0.050 & 0.385 & 0.270 & 0.043 & 0.340 & 0.230 & 0.042 & 0.330 & 0.225 & 0.043 & 0.335 & 0.235 \\
DPR & 0.060 & 0.480 & 0.340 & 0.062 & 0.495 & 0.355 & 0.056 & 0.455 & 0.315 & 0.058 & 0.470 & 0.325 & 0.057 & 0.460 & 0.320 \\  
SPECTER     & 0.043 & 0.345 & 0.252 & 0.046 & 0.363 & 0.270 & 0.039 & 0.317 & 0.221 & 0.039 & 0.313 & 0.219 & 0.039 & 0.317 & 0.228 \\
SciBERT     & 0.005 & 0.044 & 0.036 & 0.007 & 0.054 & 0.045 & 0.004 & 0.030 & 0.024 & 0.002 & 0.018 & 0.015 & 0.004 & 0.033 & 0.026\\
SciBERT-FT  & 0.055 & 0.350 & 0.262 & 0.056 & 0.362 & 0.274 & 0.051 & 0.331 & 0.243 & 0.050 & 0.327 & 0.241 & 0.050 & 0.329 & 0.242 \\
Qwen2-7B    & 0.057 & 0.442 & 0.310 & 0.059 & 0.457 & 0.327 & 0.052 & 0.415 & 0.275 & 0.053 & 0.420 & 0.287 & 0.053 & 0.419 & 0.287\\
Qwen3-7B    & 0.072 & 0.566 & 0.399 & 0.074 & 0.576 & 0.413 & 0.068 & 0.542 & 0.366 & 0.072 & 0.569 & 0.385 & 0.069 & 0.547 & 0.371\\
Qwen3-7B-FT   & 0.075 & 0.583 & \underline{0.409} & \underline{0.082} & 0.593 & \underline{0.425} & 0.070 & 0.558 & \underline{0.377} & 0.074 & 0.586 & 0.387 & 0.071 & 0.563 & \underline{0.373} \\
Memtron     & 0.064 & 0.502 & 0.362 & 0.066 & 0.516 & 0.379 & 0.060 & 0.474 & 0.335 & 0.062 & 0.488 & 0.342 & 0.061 & 0.479 & 0.333 \\
GraphCL     & \underline{0.083} & \underline{0.615} & 0.396 & 0.079 & \underline{0.620} & 0.410 & \underline{0.076} & \underline{0.598} & 0.372 & \underline{0.079} & \underline{0.620} & \underline{0.388} & \underline{0.081} & \underline{0.596} & 0.372\\
AchGNN & \textbf{0.086} & \textbf{0.634} & \textbf{0.410} & \textbf{0.084} & \textbf{0.640} & \textbf{0.447} & \textbf{0.082} & \textbf{0.614} & \textbf{0.381} & \textbf{0.081} & \textbf{0.636} & \textbf{0.388} & \textbf{0.083} & \textbf{0.613} & \textbf{0.374} \\
\hdashline
AchGNN\_noMath & 0.083 & 0.619 & 0.398 & 0.081 & 0.624 & 0.431 & 0.079 & 0.600 & 0.370 & 0.078 & 0.620 & 0.376 & 0.080 & 0.598 & 0.366 \\
\bottomrule
\end{tabular}
\label{tab:evaluation}
\end{table*}

\begin{table}[t]
\centering
\caption{Overall results for the $k = 10$ retrieved documents in \textbf{PwC}. SciBERT-FT results as reported by Ostendorff et al.~\cite{Ostendorff2022a}.}
\label{tab:evaluation-pwc}
\small
\begin{tabular}{l%
ccc%
cccccccc}
\toprule
\textit{Aspects} $\longrightarrow$ & %
    \multicolumn{3}{c}{General} & %
    \multicolumn{3}{c}{Task} \\
\cmidrule(lr){2-4} %
    \cmidrule(lr){5-7}
Methods $\downarrow$ & %
    P & R & MRR & %
    P & R & MRR\\
\midrule
SciBERT-FT & %
    - & - & - & %
    \textbf{0.569} & \textbf{0.242} & \textbf{0.708}\\
AchGNN & %
    0.486 & 0.217 & 0.529 & %
    0.478 & 0.214 & 0.524\\
\midrule
    & \multicolumn{3}{c}{Method} & \multicolumn{3}{c}{Dataset}\\
    \cmidrule(lr){1-1} %
    \cmidrule(lr){2-4} %
    \cmidrule(lr){5-7}
    SciBERT-FT & %
    0.407 & 0.168 & \textbf{0.588} & \textbf{0.270} & 0.374 & \textbf{0.533}\\
AchGNN & %
    \textbf{0.512} & \textbf{0.236} & 0.563 & 0.201 & \textbf{0.469} & 0.501\\
    \bottomrule
\end{tabular}
\end{table}

\subsection{Evaluated models}\label{sec:evaluatedmodels}
We compare AchGNN against a set of diverse state-of-the-art approaches consisting of embedding-based and graph-based baselines.
\textbf{Approach0}~\cite{zhong2022applying}, a formula similarity search engine, is the top-performing system in the formula search category, whereas \textbf{DPR} (Dense Passage Retriever)~\cite{reusch2022transformer}, a neural retriever trained for mathematical formula similarity, is the best-performing LLM-based embedding model in the text-based search category in recent MathIR evaluation tasks~\cite{mansouri2022advancing}.
\textbf{Mabowdor}~\cite{23mabowdor}, a Math-Aware Best-of-Worlds Domain-Optimized Retriever, is the best-performing hybrid approach on the ArqMATH dataset. 
It combines an unsupervised structure search for similar math formulae with a dense retriever leveraging LLMs for text content similarity, integrating both into a combined score ranker.
\textbf{SPECTER}~\cite{2020-specter}, and \textbf{SciBERT}~\cite{2019SciBERT} are both BERT-based language models that have been effectively used in \textit{aspect}-based CbRPR and often serve for robust baseline evaluations~\cite{2020OstendorffWiki,2025AspectCodes}.
The \textbf{SciBERT-FT}~\cite{Ostendorff2022a} variant employs a fine-tuned Siamese architecture based on SciBERT and currently represents the best-performing method for \textit{aspect}-based CbRPR on the PwC dataset~\cite{2020PapWitCodeData}.
Since LLM embeddings are predominant in CbRPR, we select state-of-the-art LLM embedding models from MTEB~\cite{muennighoff-etal-2023-mteb}.
The MTEB leaderboard ranking\footnote{\url{https://huggingface.co/spaces/mteb/leaderboard}} lists \textbf{Qwen2-7B}~\cite{qwen2025qwen25technicalreport} as the model with the highest overall performance as of October 2025.
The MTEB benchmark further includes domain-specific tasks. Notably, \textbf{Qwen3-7B}~\cite{qwen3embedding} attains state-of-the-art performance on the arXiv clustering task, with a majority of mathematics and physics research papers.
Building on this, \textbf{Qwen3-7B-FT} denotes a fine-tuned variant trained using instruction-tuned samples and contrastive-loss~\cite{shao2024deepseekmath,lee2025nvembedimpr}.
Another MTEB subtask involves clustering of question–answer pairs from Math Stack Exchange. On this task, \textbf{Memtron}~\cite{2025memtron} demonstrates the best performance among all evaluated models.
Lastly, Graph Contrastive Learning (\textbf{GraphCL})~\cite{you2020graphCL} is a top-performing approach~\cite{21SIGIRgrCL,cai2023lightgcl} leveraging a joint graph for recommendation, making it a strong graph-based baseline for our evaluation.
All models contain fewer than 8B parameters, reflecting realistic deployment constraints in large-scale digital libraries.

\subsection{Evaluation \& Training Setup}
We evaluate all models under an \textit{aspect}-specific retrieval paradigm, where recommendations are generated and assessed independently for each \textit{aspect}.
Let $\mathcal{AS} = \{as_1, \dots, as_n\}$ denote the set of $n$ \textit{aspects} defined in the dataset.
For each \textit{aspect} $as_j \in \mathcal{AS}$, we construct independent training and test splits, resulting in $n$ \textit{aspect}-specific retrieval tasks. 
During evaluation, each document in the test split corresponding to \textit{aspect} $a_j$ is treated as a query seed. 
Given a seed document $d_s$, the model computes an \textit{aspect}-specific representation $\vec{d}_s(a_j)$, which is used to retrieve the top-$k$ candidate documents via $k$-nearest neighbor search.
Document similarity is measured using cosine similarity, following standard practice in \textit{aspect}-based CbRPR~\cite{pinedo2025RSSurvey,2025ZhangCiteIntentSurvey}.
In addition to \textit{aspect}-wise evaluation, we report results under a \textit{general} setting.
Here, all test documents from all \textit{aspects} are pooled into a single test set, and retrieval is performed without specifying any \textit{aspect}. 
This evaluation measures overall recommendation effectiveness and assesses whether performance gains observed in \textit{aspect}-specific scenarios generalize to an \textit{aspect}-agnostic retrieval setting.

Trained models (SciBERT-FT, Qwen3-7B-FT, GraphCL, and Ach-GNN) are provided with \textit{aspect} information, whereas the remaining base models produce \textit{aspect}-agnostic representations.
To prevent models from exploiting abstract reference cues (such as in Figure~\ref{fig:exampleSILVER}), we remove all explicit mentions of other zbMATH Open documents from abstracts prior to training and evaluation.
Due to its expert-curated nature and limited size, GoldRiM is used exclusively for testing.
SilverRiM and PwC are each split into 75\% training and 25\% test sets, with 5-fold cross-validation employed to ensure robustness.
For training, SilverRiM recommendation pairs are used as positive samples.
Hard negatives are constructed by retrieving an equal number of top-ranked candidates using Qwen3-7B embeddings with cosine similarity, excluding documents that appear in the training data.

We report Precision@k (P), Recall@k (R), and Mean Reciprocal Rank (MRR). 
While the average number of GoldRiM and SilverRiM recommendations per query is 5.25 and 2.50, respectively, we set $k = 10$ to evaluate ranking quality beyond minimal recall and to remain comparable with prior work on PwC.
The retrieval corpus consists of 3.8 million English-language zbMATH Open documents.
For AchGNN, the number of paper neighbors was set to 15, author neighbors to five, the number of layers $L=2$, and $\lambda=0.2$.

\subsection{Experimental results}
We report mean values across the 5 cross-validation folds in Table~\ref{tab:evaluation}; the corresponding standard deviations range from 0.005 to 0.021 across all metrics and models, indicating stable performance and consistent behavior across folds.
Although SPECTER outperforms SciBERT among base models, its gains do not translate into improved fine-tuning outcomes for mathematical CbRPR; this behavior has also been observed in prior state-of-the-art work, where fine-tuned variants of SPECTER underperform relative to SciBERT~\cite{Ostendorff2022a}.
SciBERT-FT, fine-tuned on SilverRiM, performs significantly better but still falls short compared to the other approaches.
This is particularly noteworthy since SciBERT-FT is the best performing model on the PwC task~\cite{Ostendorff2022a,2025AspectCodes}.
MathIR methods (Approach0, DPR, and Mabowdor), which perform strongly on recent MathIR evaluation tasks~\cite{mansouri2022advancing}, exhibit underperformance in CbRPR (Approach0 shows slight improvements but still lags behind recent fine-tuned LLM embeddings).
This indicates that general-purpose embedding models outperform purely formula-centric MathIR approaches in capturing CbRPR, even when the latter are fine-tuned on mathematical content.
However, incorporating specialized formula representations into LLMs remains a promising direction for future exploration~\cite{10.1109/TLT.2023.3333669}.
Table~\ref{tab:evaluation-pwc} compares AchGNN on the PwC dataset and its three \textit{aspects} with the reported results of SciBERT-FT~\cite{Ostendorff2022a} (here fine-tuned on PwC).
The relative competitive performance of AchGNN on PwC indicates a more robust approach across multiple datasets. 
This is promising, since a general lack of generalizability across datasets of recommender systems is rather common~\cite{McElfreshKV0W22,ChinCC22} and also evident from our evaluation of the baselines on our math-specific datasets.

Among base LLM embedding models, Qwen3-7B achieves the strongest performance. 
This result is consistent with its strong performance on math-intensive clustering tasks in MTEB, suggesting improved representations of formal and abstract text. 
The fine-tuned variant, Qwen3-7B-FT, further improves performance over the base model. 
However, its performance still falls short of graph-based approaches. 
This suggests that simple fine-tuning alone is insufficient to capture implicit CbRPR in mathematics, and that performance bottlenecks in this domain are not solely due to limited supervision~\cite{weller2025theoretical}.

GraphCL outperforms embedding-only baselines across most \textit{aspects} and both datasets, highlighting the importance of structural signals derived from a joint paper-author graph.
AchGNN achieves the best performance across all \textit{aspects}, metrics, and datasets (GoldRiM and SilverRiM), with a \textit{General} MRR of 0.593 (on GoldRiM).
This corresponds to a 26\% relative improvement over GraphCL, despite operating on the same graph structure, and more than a five-fold improvement over SciBERT-FT.
Addressing RQ2, AchGNN’s modeling of \textit{aspects} during representation learning captures representations for mathematical documents more effectively than the \textit{aspect}-agnostic contrastive learning employed by GraphCL.
Lastly, the performance changes between evaluations on SilverRiM and GoldRiM across all models are consistent, mostly preserving the same model order (see the bar chart representations of Table~\ref{tab:evaluation} in our repository).
This indicates that GoldRiM, despite its inherent bias from a single annotator and limited scope, can still be used as a meaningful evaluation dataset.
On SilverRiM, we observe a drop in precision due to the smaller number of relevant recommendations per seed (See Table~\ref{tab:dataset_characteristics}), which limits the achievable P@k, while recall improves; however, the comparatively lower MRR and the reduced margin over competing approaches suggest that the noisier and less curated nature of SilverRiM may dilute ranking quality, an effect that needs further investigation.

\subsection{Ablation Study}\label{sec:ablation}
We conduct an ablation study to quantify the contribution of individual \textit{aspects} and graph signals in AchGNN.
Each variant is trained by removing one supervision source while keeping the architecture, optimization, and evaluation protocol fixed.
We further analyze the authorship contributions added to the graph structure and evaluate performance across varying GNN layer parameters and the number of neighborhood nodes.

\subsubsection{Aspect Removal}
\begin{figure}[t]
    \centering
    \includegraphics[width=0.38\textwidth]{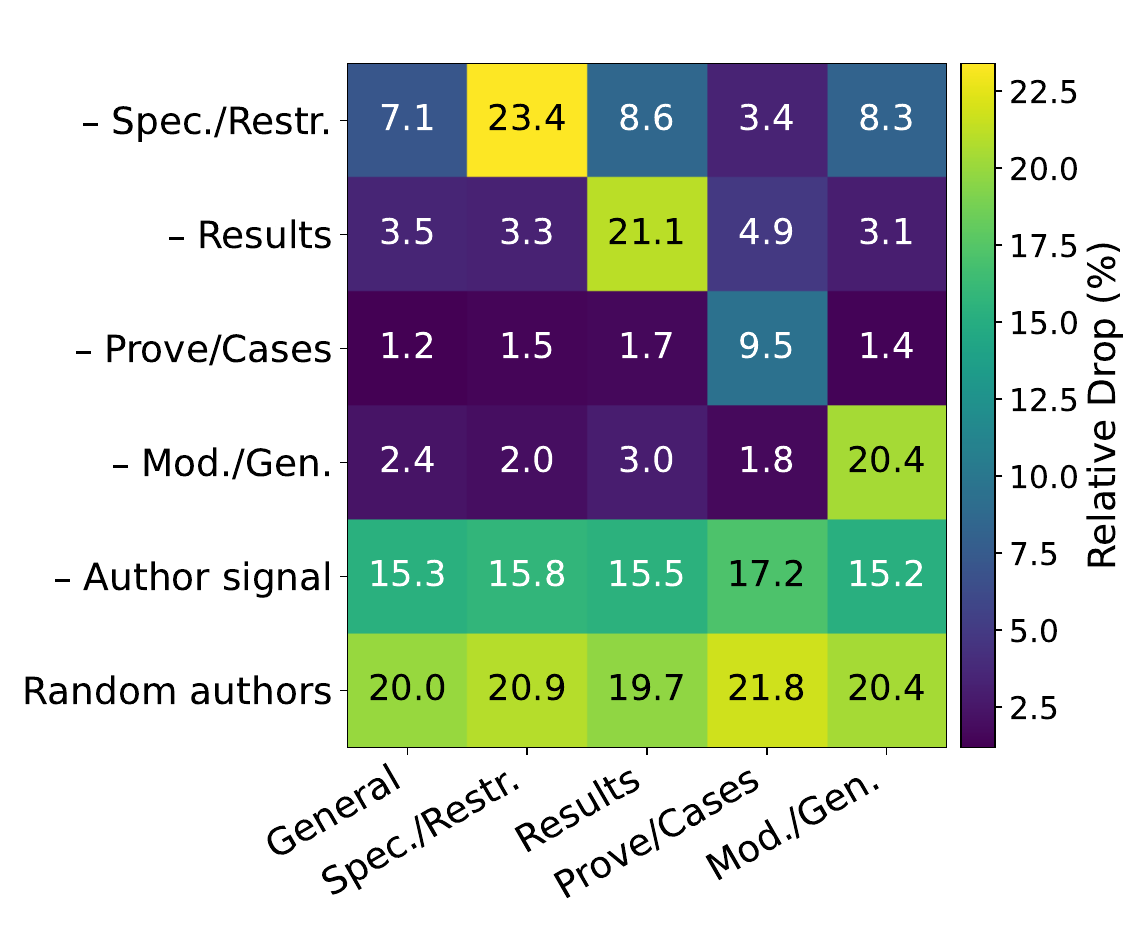}
    \caption{Relative R@10 performance drops for removing and modifying \textit{aspects} (y-axis) on GoldRiM.}
    \label{fig:randAuthor}
\end{figure}

Figure~\ref{fig:randAuthor} summarizes the relative performance drops on Recall@10.
Removing \textit{Specialization/Restriction} supervision results in the greatest performance degradation.
In contrast, removing \textit{Prove / Cases} results in the smallest performance drop, reflecting its limited training size (see Figure~\ref{fig:aspect_labels}) and highly localized relevance patterns.
\textit{Aspect} ablations primarily degrade performance on the removed \textit{aspect}, with limited spillover to others.
This is further underlined by the relatively small performance drops on \textit{General}.
This indicates that AchGNN does not collapse multiple relevance dimensions into a single embedding.
Instead, \textit{aspect}-conditioned scoring successfully disentangles relevance criteria~\cite{ZHANG2023Magnn}.
We further normalized the performance degradation against the number of removed edges during training but found the same relative performance changes with respect to each \textit{aspect}.

\subsubsection{Math Formula Removal}
AchGNN\_noMath (Table~\ref{tab:evaluation}) denotes a variant of AchGNN in which all mathematical formulae are removed from text input prior to embedding generation.
This allows us to isolate formulae from text and assess whether performance gains decrease with formulae removal and the extent to which excluding formulae from text affects embeddings, since they are universal.
The results indicate that, although mathematical formulae contribute a useful signal, the majority of the model’s performance is driven by textual and structural information captured in the graph. 
In particular, the relatively minor impact of formula removal suggests that current embedding models already encode a substantial amount of semantic information from surrounding natural language, and that formulae in their raw form provide only incremental benefits.

\subsubsection{Validation of Authorship Lineage}\label{sec:ablation-randAuth}
To validate the role of authorship lineage, we perform authorship ablation and study the effects of adding purposefully generated noise to the data by evaluating randomized authorship.
Both variants cause substantial degradation, often even greatly exceeding that of any single-\textit{aspect} removal.
This is further apparent in the normalized performance drop, since the authorship signal accounts for only about 12\% of the connections in the graph, which is on par with the smallest \textit{aspect} set, \textit{Prove / Cases}.
Randomized authorship performs worst and significantly degrades performance even compared to removing authorship information entirely in absolute terms. 
This strongly indicates that authorship information contributes to semantic lineage rather than mere added graph connectivity~\cite{hulek2023mathAuthor}.
Furthermore, it highlights that AchGNN leverages meaningful intellectual structure beyond textual similarity.

\subsubsection{Effect of Neighborhood Sampling Size and GNN Depth}\label{sec:ablation-neighborhood}
\begin{figure}[t]
  \centering
  \includegraphics[width=0.4\textwidth]{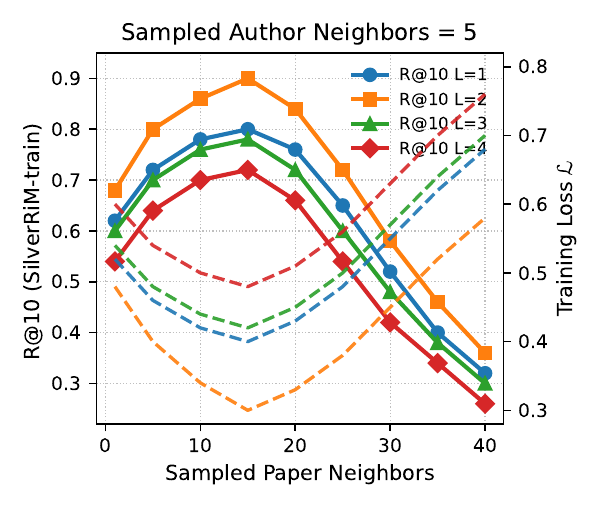}
  \caption{Effects of sampled neighborhood sizes and GNN depths ($L$) on SilverRiM training, where sampled papers refers to the number of neighbors sampled at each layer. 
  }
  \label{fig:gnn_bias_variance_silverrim}
\end{figure}

Figure~\ref{fig:gnn_bias_variance_silverrim} evaluates the sensitivity of AchGNN to paper-node neighborhood sampling and GNN depth on SilverRiM, with the author-node neighborhood size fixed to five.
AchGNN performs best with moderate neighborhood sizes, achieving its lowest training loss and highest Recall@10 with 15 paper neighbors, 5 author neighbors, and $L=2$ GNN layers.
Increasing the paper neighborhood size initially improves performance.
This confirms the benefit of incorporating citation and authorship context beyond single node representation through embedding~\cite{2017GraphSage,zhong2023hierarchical,2022GraphCitRec}.
Further increases lead to degraded Recall@10 and higher loss due to over-smoothing~\cite{you2020graphCL,cai2020note}.
For author nodes, sampling a single neighbor reduces performance (over 5\%).
In contrast, increasing the cap from five to ten yields only marginal changes (below 1\%).
This behavior aligns with the SilverRiM average author-degree distribution, which is approximately 2.5 authors, causing author neighborhoods to saturate once moderate sampling thresholds are reached.
Overall, AchGNN benefits most from moderate neighborhood sizes, balancing structural context aggregation with over-smoothing avoidance.

\section{Conclusion and Future Work}
This work addresses a fundamental gap in content-based research paper recommendation (CbRPR) in mathematics, a domain in which relatedness is rarely captured by surface-level similarity. 
Through an expert-driven analysis, we demonstrate that mathematical CbRPR is inherently \textit{aspect}-driven, grounded in conceptual relations such as theorem generalization, dual formulations, and proof reuse, which remain invisible to state-of-the-art CbRPR approaches.
To support a systematic study of this problem, we introduced GoldRiM, a high-quality test dataset designed to assess and uncover mathematical relevance, and SilverRiM, a scaled version that enables training and evaluation of \textit{aspect}-aware CbRPR in mathematics.
Motivated by the limitations of standalone LLM embeddings for mathematical content, we proposed AchGNN, an \textit{aspect}-conditioned heterogeneous graph neural network that jointly models textual semantics, citation structure, and author identity. 
Experimental results show that AchGNN consistently outperforms state-of-the-art \textit{aspect}-based CbRPR methods.
Beyond quantitative improvements, our evaluation demonstrated that effective mathematical CbRPR requires structural and relational modeling rather than purely semantic. 
Ablation analyses confirm that AchGNN’s improvements stem from \textit{aspect} supervision, authorship lineage, and selected neighborhood and depth configurations.
We publicly released our datasets and code: \repolink{}.

Hosting recommendations on the MaRDI platform enables us to evaluate user-based interactions in the future, allowing us to explore state-of-the-art approaches for user-item interaction graphs and hybrid recommendation settings that combine content-based and collaborative signals.
Furthermore, we plan to validate identified math-specific \textit{aspects} with additional reviewers across larger subfields.
This will help refine the aspect taxonomy and determine whether additional aspects—such as methodological similarity or historical influence—should be incorporated.
This will help refine the aspect taxonomy and determine whether additional aspects should be incorporated.
In addition, the current graph size is limited to zbMATH Open. 
zbMATH Open articles, however, often refer to external articles, most notably those on arXiv.
Expanding the network of paper relationships and authorship lineage across multiple datasets can potentially improve generalizability and position AchGNN as a standard for mathematical research recommender systems.

\section*{Acknowledgments}
Funded by the Deutsche Forschungsgemeinschaft (DFG, German Research Foundation) –  \href{https://gepris.dfg.de/gepris/projekt/437179652}{437179652}; \href{https://gepris.dfg.de/gepris/projekt/567156310}{567156310};
\href{https://gepris.dfg.de/gepris/projekt/554559555}{554559555}.
This work was supported by the Lower Saxony Ministry of Science and Culture and the VW Foundation.
The authors gratefully acknowledge the computing time granted by the KISSKI project. 
The calculations for this research were conducted with computing resources under the project \textit{MathRecSys}.
We utilized AI models to proofread, enhance grammar, and improve
sentence clarity. Every sentence in the resulting text is checked by
the authors, and citations are provided wherever available. We take
full responsibility for the text in this manuscript.

\bibliographystyle{ACM-Reference-Format}
\bibliography{main_}

\appendix

\makeAIUsageCard

\clearpage
\onecolumn
\hypertarget{annotation}{}
\pagestyle{empty}
\lstset{
  basicstyle=\footnotesize\ttfamily,
  breaklines=true,
  breakatwhitespace=false,
  columns=flexible,
  numbers=none
}

\definecolor{Primary}{RGB}{59, 130, 246}    
\definecolor{PrimaryDark}{RGB}{30, 64, 175} 
\definecolor{LightBg}{RGB}{239, 246, 255}   
\definecolor{TextDark}{RGB}{31, 41, 55}     
\definecolor{TextMuted}{RGB}{107, 114, 128} 

\begin{tikzpicture}[remember picture, overlay]
  \fill[Primary] ([xshift=0cm,yshift=0cm]current page.north west) rectangle ([xshift=\paperwidth,yshift=-0.4cm]current page.north west);
\end{tikzpicture}

\vspace{0.8cm}
\begin{center}
  {\fontsize{22}{26}\selectfont\sffamily\bfseries \textcolor{PrimaryDark}{CiteAssist}}\\[0.2em]
  {\Large\sffamily\scshape \textcolor{TextMuted}{Citation Sheet}}\\[0.8em]
  {\small\sffamily Generated with \href{https://citeassist.uni-goettingen.de/}{\textcolor{Primary}{\texttt{citeassist.uni-goettingen.de}}}
  \CiteAssistCite{}
  }\end{center}

\begin{center}
\vspace{1em}
\begin{tikzpicture}
\draw[Primary, line width=0.6pt] (0,0) -- (\textwidth,0);
\end{tikzpicture}
\vspace{1.2em}
\end{center}

\begin{tcolorbox}[enhanced,
                 frame hidden,
                 boxrule=0pt,
                 borderline west={2pt}{0pt}{Primary},
                 colback=LightBg,
                 sharp corners,
                 breakable,
                 fonttitle=\sffamily\bfseries\large,
                 coltitle=Primary,
                 title=BibTeX Entry,
                 attach title to upper={\vspace{0.2em}\par},
                 left=12pt]
\lstset{
    inputencoding = utf8,  
    extendedchars = true,  
    literate      =        
      {á}{{\'a}}1  {é}{{\'e}}1  {í}{{\'i}}1 {ó}{{\'o}}1  {ú}{{\'u}}1
      {Á}{{\'A}}1  {É}{{\'E}}1  {Í}{{\'I}}1 {Ó}{{\'O}}1  {Ú}{{\'U}}1
      {à}{{\`a}}1  {è}{{\`e}}1  {ì}{{\`i}}1 {ò}{{\`o}}1  {ù}{{\`u}}1
      {À}{{\`A}}1  {È}{{\`E}}1  {Ì}{{\`I}}1 {Ò}{{\`O}}1  {Ù}{{\`U}}1
      {ä}{{\"a}}1  {ë}{{\"e}}1  {ï}{{\"i}}1 {ö}{{\"o}}1  {ü}{{\"u}}1
      {Ä}{{\"A}}1  {Ë}{{\"E}}1  {Ï}{{\"I}}1 {Ö}{{\"O}}1  {Ü}{{\"U}}1
      {â}{{\^a}}1  {ê}{{\^e}}1  {î}{{\^i}}1 {ô}{{\^o}}1  {û}{{\^u}}1
      {Â}{{\^A}}1  {Ê}{{\^E}}1  {Î}{{\^I}}1 {Ô}{{\^O}}1  {Û}{{\^U}}1
      {œ}{{\oe}}1  {Œ}{{\OE}}1  {æ}{{\ae}}1 {Æ}{{\AE}}1  {ß}{{\ss}}1
      {ẞ}{{\SS}}1  {ç}{{\c{c}}}1 {Ç}{{\c{C}}}1 {ø}{{\o}}1  {Ø}{{\O}}1
      {å}{{\aa}}1  {Å}{{\AA}}1  {ã}{{\~a}}1  {õ}{{\~o}}1 {Ã}{{\~A}}1
      {Õ}{{\~O}}1  {ñ}{{\~n}}1  {Ñ}{{\~N}}1  {¿}{{?\`}}1  {¡}{{!\`}}1
      {„}{\quotedblbase}1 {“}{\textquotedblleft}1 {–}{$-$}1
      {°}{{\textdegree}}1 {º}{{\textordmasculine}}1 {ª}{{\textordfeminine}}1
      {£}{{\pounds}}1  {©}{{\copyright}}1  {®}{{\textregistered}}1
      {«}{{\guillemotleft}}1  {»}{{\guillemotright}}1  {Ð}{{\DH}}1  {ð}{{\dh}}1
      {Ý}{{\'Y}}1    {ý}{{\'y}}1    {Þ}{{\TH}}1    {þ}{{\th}}1    {Ă}{{\u{A}}}1
      {ă}{{\u{a}}}1  {Ą}{{\k{A}}}1  {ą}{{\k{a}}}1  {Ć}{{\'C}}1    {ć}{{\'c}}1
      {Č}{{\v{C}}}1  {č}{{\v{c}}}1  {Ď}{{\v{D}}}1  {ď}{{\v{d}}}1  {Đ}{{\DJ}}1
      {đ}{{\dj}}1    {Ė}{{\.{E}}}1  {ė}{{\.{e}}}1  {Ę}{{\k{E}}}1  {ę}{{\k{e}}}1
      {Ě}{{\v{E}}}1  {ě}{{\v{e}}}1  {Ğ}{{\u{G}}}1  {ğ}{{\u{g}}}1  {Ĩ}{{\~I}}1
      {ĩ}{{\~\i}}1   {Į}{{\k{I}}}1  {į}{{\k{i}}}1  {İ}{{\.{I}}}1  {ı}{{\i}}1
      {Ĺ}{{\'L}}1    {ĺ}{{\'l}}1    {Ľ}{{\v{L}}}1  {ľ}{{\v{l}}}1  {Ł}{{\L{}}}1
      {ł}{{\l{}}}1   {Ń}{{\'N}}1    {ń}{{\'n}}1    {Ň}{{\v{N}}}1  {ň}{{\v{n}}}1
      {Ő}{{\H{O}}}1  {ő}{{\H{o}}}1  {Ŕ}{{\'{R}}}1  {ŕ}{{\'{r}}}1  {Ř}{{\v{R}}}1
      {ř}{{\v{r}}}1  {Ś}{{\'S}}1    {ś}{{\'s}}1    {Ş}{{\c{S}}}1  {ş}{{\c{s}}}1
      {Š}{{\v{S}}}1  {š}{{\v{s}}}1  {Ť}{{\v{T}}}1  {ť}{{\v{t}}}1  {Ũ}{{\~U}}1
      {ũ}{{\~u}}1    {Ū}{{\={U}}}1  {ū}{{\={u}}}1  {Ů}{{\r{U}}}1  {ů}{{\r{u}}}1
      {Ű}{{\H{U}}}1  {ű}{{\H{u}}}1  {Ų}{{\k{U}}}1  {ų}{{\k{u}}}1  {Ź}{{\'Z}}1
      {ź}{{\'z}}1    {Ż}{{\.Z}}1    {ż}{{\.z}}1    {Ž}{{\v{Z}}}1  {ž}{{\v{z}}}1
  }
\begin{lstlisting}
@inproceedings{Satpute2026,
  author={Satpute, Ankit and Greiner-Petter, Andre and Giessing, Noah and Teschke, Olaf and Schubotz, Moritz and Aizawa, Akiko and Gipp, Bela},
  booktitle={Proceedings of 49th International ACM SIGIR Conference on Research and Development in Information Retrieval (SIGIR 26)},
  month={July.},
  publisher={ACM},
  title={Aspect-Aware Content-Based Recommendations for Mathematical Research Papers},
  year={2026},
  address={Melbourne | Naarm, Australia},
  doi={10.1145/3805712.3809531}
}
\end{lstlisting}
\end{tcolorbox}

\vfill
\begin{tikzpicture}
\draw[Primary!40, line width=0.4pt] (0,0) -- (\textwidth,0);
\end{tikzpicture}
\begin{center}
\small\sffamily\textcolor{TextMuted}{Generated \today}
\end{center}

\end{document}